\documentclass[intlimits,twoside,a4paper]{article}

\usepackage[utf8]{inputenc}
\usepackage{graphicx}
\usepackage[eqsecnum]{cmpj3}
\usepackage{hyperref}
\issue{2026}{29}{3}{33703}
\doinumber{10.5488/CMP.29.33703}

\title[Computational investigation of a perovskite LaBiO$_{3}$]%
{Computational investigation of a perovskite LaBiO$_{3}$ for
photovoltaic, thermoelectric, and optoelectronic applications}
\author[M. M. Woldemariam et al.]{%
	M. M. Woldemariam\orcid{0000-0001-8541-6707}\thanks{Corresponding author: \email{menberu.mengesha@ju.edu.et}}, 
	N. G. Debelo\orcid{0000-0002-4911-2455}, 
	T.~K.~Tufa\orcid{0009-0008-5406-0224}, 
	E.~M.~Gurmesa\orcid{0009-0006-8738-1556}, 
	S.~H.~Didu\orcid{0000-0003-3843-8326}, 
	S.~N.~Asfaw\orcid{0000-0001-9762-0123},  
	D.~T.~Keno\orcid{0000-0002-1413-4881}%
}

\address{
	Department of Physics, Jimma University, P.O. Box 378, Jimma, Ethiopia
}

\date{Received 06 February 2026; revised 04 June 2026; accepted 11 June 2026; published 28 September 2026}

\begin{document}

\maketitle

\begin{abstract}
Using density functional theory (DFT) with the ONCVVPSP pseudopotential and PBE functional, this study investigates the structural, electronic, elastic, optical, and thermoelectric properties of the trigonal LaBiO$_{3}$ perovskite oxide (space group R3c). Ground-state parameters lattice constant, volume, bulk modulus, and its pressure derivative were determined using the equation of state. Applying the Hubbard correction (GGA+U) revealed an indirect, wide band gap of 3.51 eV. Mechanical properties, including the anisotropy factor, elastic modulus, and Poisson’s ratio, were calculated via the Voigt--Reuss--Hill averaging scheme. The bulk-to-shear modulus ratio identifies the trigonal phase as ductile. Additionally, Debye temperatures and sound velocities were computed. Optical characteristics (absorption coefficient, refractive index, and electron energy loss function) were evaluated across a 0--35 eV spectral range. Finally, semi-classical transport coefficients, including electrical conductivity, Seebeck coefficient, and power factor, were calculated to assess the material's thermoelectric potential.
\end{abstract}
\section{Introduction}
Today, an increase in energy demand all over the world has prompted
researchers to seek alternative materials that support clean and
sustainable energy. Perovskite based materials have been established
as a family of materials for light harvesting, energy conversion and
storage \cite{a1, a2, a3}. In recent decades, inorganic-organic or
organometal halide perovskites have gained prominence, although their origins trace
back to the early 20th century. For the first time, the synthesis
and physical properties of organometal lead halide perovskite was
reported in 1970 \cite{a4}. Following this, the organometal lead
halide perovskites have  emerged as a candidate for solar cell
absorbers with the first solar cell reported in 2009 \cite{a5}.
Recently, the perovskite solar cells have attracted researchers due
to a significant increase in power conversion efficiency from 3.8\% to 25.5\% \cite{a6}.

Perovskite devices use light-absorbing materials with the perovskite
crystal structure, generally expressed as ABX$_{3}$. In this
structure, A and B are cations of different sizes, while X is
typically an oxide or halide anion linking them \cite{a7, a8}. The
A-site cations, usually larger and lower in valence such as Ca or La
occupy 12-fold oxygen-coordinated positions. Smaller B-site cations,
including Co or Cr, sit in six-fold octahedral coordination. Both
sites can be modified by substituting metals or semi-metals. When
lower-valence cations replace A or B, the charge imbalance creates
oxygen vacancies. These vacancies enable mixed ionic-electronic
conductivity, mainly governed by transition-metal centers, and help
maintain the charge neutrality within the lattice \cite{a9, a10}.

Despite the growing interest in organic halide perovskite materials,
significant challenges remain before they can be broadly adopted for
large-scale commercial applications. One major concern is the
presence of toxic lead (Pb), which poses environmental and health
risks. Additionally, these materials are highly sensitive to
moisture, leading to rapid degradation \cite{a11}, and they suffer
from ion migration, which raises concerns about long-term stability
 \cite{a12}. Their ultralow thermal conductivity also contributes to
potential heating and mechanical stress during operation \cite{a13}.
Whereas ongoing research is making progress in addressing some of
these issues, others may be inherent to the nature of organic lead
halide perovskites and therefore difficult to overcome.

Perovskite oxides have garnered a considerable scientific attention
because of their diverse and flexible potential applications. These
materials are being explored across various applications, including
fuel cells, power generation, magnetic sensors, spintronics,
actuators, high-pressure magnetic and optical sensors,
supercapacitors, high-strength composites, radiation tracers,
piezoelectric  and thermoelectric material. This broad applicability
has driven extensive research efforts aimed at unlocking their full
potential \cite{a14, a15}.

To our understanding, the structural, electronic, elastic, optical,
and thermoelectric properties of LaBiO$_{3}$ have not been
thoroughly investigated using first-principles methods. We assessed
their structural stability using the commonly employed octahedral
and Goldschmidt tolerance factors. The electronic properties are
calculated by considering GGA-PBE \cite{a17}  as well as density
functional theory (DFT) with the Hubbard functional (GGA+U)
\cite{a18} for exchange correlation potential using Quantum Espresso
Package (QE). The thermodynamic properties are calculated using the
thermo-pw software interfaced with Quantum Espresso \cite{a22}.
Moreover, the optical properties were evaluated using time-dependent density functional perturbation
theory (TD-DFPT) within the independent-particle formalism and the
random phase approximation.
In addition, the semi-classical transport coefficients are
determined within the Boltzmann transport theory using BoltzTrap code
\cite{a23}. Overall, with such predictable properties, these
materials hold a considerable potential for thermoelectric and
optoelectronic applications, albeit requiring experimental
confirmation.

\section{Computational details}
In this work, the first principle computations were performed within
Quantum Espresso software Package (QE) \cite{a16} based on  the
density functional theory (DFT) as implemented within the
generalized gradient approximation (GGA) with the Hubbard correction
(GGA+U) \cite{a18}. The effective Hubbard parameter ($U_{\rm eff}$) was
calculated iteratively for La-$5d$ orbital using a linear response
formalism utilizing  DFPT with ortho-atomic projection
method~\cite{a19} and found $U_{\rm eff}=1.65$  for La ($d$). The optimized
norm-conserving Vanderbilt pseudopotential (ONCVPSP) with
Perdew-Burke-Ernzerhof (PBE) exchange-correlation functional  were
used to treat the interaction of the electrons with the ion cores as
in \cite{a19}. For the calculations, the relevant valence electrons
are La-[Xe]$5d^{1} 6s^{1}$, Bi-[Xe]$  4f^{14}  5d^{10} 6s^{2}
6p^{2}$, O-[He]$ 2s^{2} 2p^{4}$. Careful tests were performed to
ensure the convergence of calculations concerning the plane-wave
cut-off and $k$-point mesh. Crystal structure optimization is  done
using a plane wave cutoff energy of 60 Ry and the Brillouin zone
with an $8 \times 8 \times 8$ Monkhorst-Pack $k$-point grid \cite{a20}
based on the convergence criteria --- energy $10 ^{-5}$~Ry, force
$10 ^{-3}$ Ry/Bohr, and cell pressure 0.5~kbar --- with the help of
Brodyden--Fletcher--Goldfarb--Shanno (BFGS) method \cite{a21}.
Thermodynamic properties were calculated using   Thermo-pw code
\cite{a22}. The semi-classical transport coefficients were obtained
by converting the Quantum Espresso output generated with a dense
$k$-point mesh into a format compatible with the BoltzTraP code
\cite{a23}.

\section{Result and discussions}
\subsection{Structure properties}
\begin{figure}[h]
	\centerline{\includegraphics[scale=0.21]{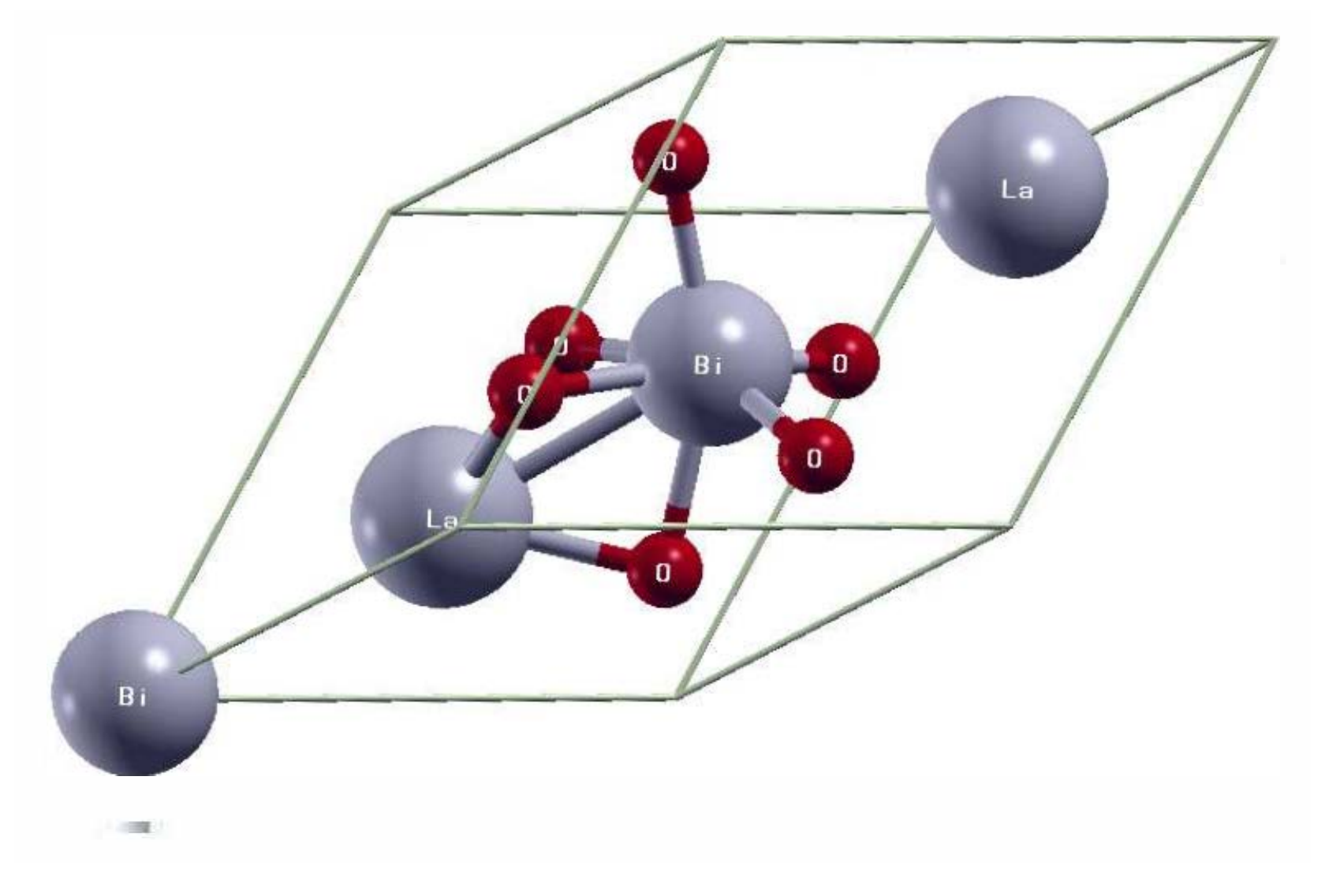}}
	\caption{(Colour online) Crystallographic structures of   R3c  LaBiO$_{3}$.}
	\label{fig-1}
\end{figure}

In this study, the trigonal structure of LaBiO$_{3}$ [with space
group R3c (161)], was considered as shown in figure~\ref{fig-1}. In
the trigonal structure of LaBiO$_{3}$, both La$^{3+}$ and Bi$^{3+}$
cations are coordinated to six oxygen atoms forming distorted
octahedra. Unlike the corner sharing network found in cubic
perovskites, these LaO$_{6}$ and BiO$_{6}$ units are aranged in a
more densely packed framework characterized by edge sharing and face
sharing connectivity. Especially, the Bi$^{3+}$ ion is coordinated to
six equivalent O$^{2-}$ atoms, resulting in a distorted BiO$_{6}$
octahedron  where the distortion is derived by the streochemical
activity of the bismuth $6s^{2}$ lone pair. The La$^{3+}$ ion is
surrounded by six equivalent O$^{2-}$ atoms, where the La--O bonds
measure 2.43 $\text{\AA}$, the Bi--O bond length is 2.11 $\text{\AA}$ and the La--Bi measure
3.2 $\text{\AA}$.

The Goldschmidt factor is a crucial parameter for providing
structural information on perovskites. The modified form of
Goldschmidt tolerance factor $(\tau)$ is given by \cite{a24}:
\begin{equation}
 \tau = \frac{1}{\sqrt{2}} ~ \frac{(r_{\rm La} + r_{\rm O})}{(r_{\rm Bi} + r_{\rm O})},
\end{equation}
where $r_{\rm La}$, $r_{\rm Bi}$ and $r_{\rm O}$ are ionic radii, respectively.

Typically, a tolerance factor of 0.9--1.0 produces an ideal cubic
arrangement, while a factor between $0.71$ and $0.9$ induces a
distorted perovskite structure featuring tilted octahedra.
Non-perovskite structures are formed when the tolerance factor is
higher $(>1)$ or lower $(<0.71)$ \cite{a24}. La$^{3+}$ exhibits a
calculated tolerance factor of $0.81$, which verifies the stability of a cubic
perovskite structure, as displayed in table~\ref{Tab1}.

\begin{table}[htb]
\caption{The ionic radii and calculated tolerance factor of
LaBiO$_3$.}
\label{Tab1}
%\vspace{2ex}
\begin{center}
\renewcommand{\arraystretch}{1.1}
\begin{tabular}{c c c c}
\hline
Ions &  Ionic radius ($\text{\AA}$) & Tolerance factor \\
\hline
La$^{+3}$    &   1.36  &       \\
\hline
Bi$^{+3}$   &   1.02  &   0.81 \\
\hline
O$^{-2}$    &   1.4  &        \\
\hline
\end{tabular}
\renewcommand{\arraystretch}{1}
\end{center}
\end{table}

\begin{figure}[htb]
\centerline{\includegraphics[scale=0.5]{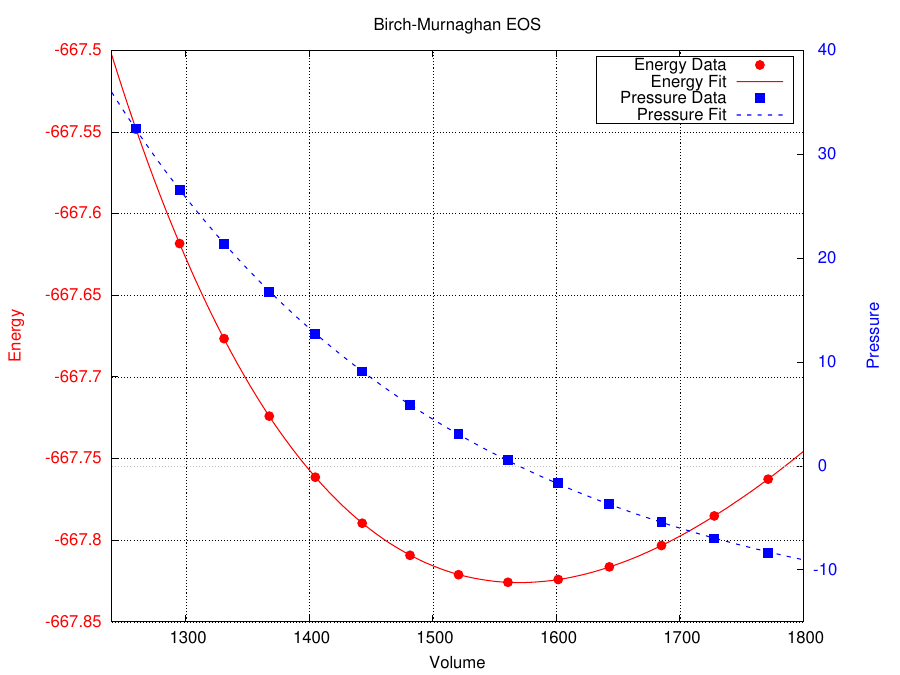}}
\caption{(Colour online) Volume versus energy and pressure.} \label{fig-2}
\end{figure}

In this study, the structural stability and total energy as a
function of atomic volume for La$^{3+}$ were investigated through
total energy minimization using PBE-GGA and GGA+U
exchange-correlation functionals.  From the optimized geometry, the
lattice constant is calculated as 11.3 a.u. using GGA-PBE and 11.6
a.u. using GGA+U. The equilibrium lattice potential, related to the
total energy, was evaluated via a set of strained lattice
configurations. These findings enable the calculation of the
equilibrium unit cell volume $(V_0)$, bulk modulus $(B_0)$, and its
pressure derivatives $(B_0')$. A series  of volume-dependent total
energy calculations can be fitted to an equation of state consistent
with Murnaghan equation of state (energy versus volume) \cite{a25} as 
\begin{equation}
	E(V) = E_0 + \frac{B_0 V}{B_0'} \left[ \frac{\left({V_0}/{V}\right)^{B_0'}}{B_0' - 1} + 1 \right] - \frac{B_0 V_0}{B_0' - 1},
\end{equation}
where $B_0$ is an equilibrium bulk modulus that effectively measures
the curvature of the energy versus the volume curve about the relaxed
volume $V_0$, and  $B_0^{\prime}$ is the derivative of the bulk modulus.
The calculated total energy is fitted into the Murnaghan equation
\cite{a25} for a number lattice constant demonstrated as shown
in figure~\ref{fig-2}. The computed values of the unit cell volume
($V_0)$, bulk modulus $(B_0)$, and its pressure derivative $(B_0')$
are presented in table~\ref{Tab2}. Furthermore, the results indicate that the
GGA+U approximation yields a lower bulk modulus and its derivative,
along with an increase in the lattice constants and volume, compared
to those obtained using GGA-PBE.

The formation energy of compound is among the most crucial parameters
determining its stability. The higher are the negative values of  the formation energy,
the more stable the compound is predicted to be \cite{a26}. The
formation energy, $\Delta H$,  for this system as in \cite{a27} can
be calculated as:
\begin{equation}
\Delta H ({\rm LaBiO_{3}})=\{E_{\rm tot} ({\rm LaBiO_{3}})-[N_{\rm La} E_{\rm tot} ({\rm La})+
N_{\rm Bi} E_{\rm tot}({\rm Bi})+N_{\rm O} E_{\rm tot} ({\rm O})]\}\frac {1}{N_{\rm tot}},
\end{equation}
where $N_{\rm La}$, $N_{\rm Bi}$, and $N_{\rm O}$ are total numbers of  La~$
(=1)$, Bi~$(=1)$, and O~$(=3)$ atoms respectively, and $N_{\rm tot}$ is
total number of atoms in LaBiO$_{3}$~$(=5)$. The formation energy is
determined and given in table~\ref{Tab2}. The negative values indicate that the
energy gained during the formation confirms that the compounds are
stable.

%Table 2. Computed equilibrium lattice constants, unit cell volumes, bulk moduli, their pressure derivatives, and formation energies of trigonal structure R3c (161)   LaBiO$_{3}$.

\begin{table}[htb]
	\centering % Better than \begin{center}
		\caption{Calculated values of equilibrium lattice constant, unit cell volume, bulk modulus and its derivative of LaBiO$_{3}$ in comparison with existing works.} 
		\label{Tab2}
		\vspace{1ex}
		\renewcommand{\arraystretch}{1.1}
		\begin{tabular}{l c c c c c c} % 7 columns defined
			\hline
			Source & Phase & $a$ (a.u.) & Vol (a.u.$^3$) & $B$ (GPa) & $B'$ & $\Delta H$ (eV/atom) \\
			\hline
			GGA-PBE & Fm-3m & 11.3 & 1442.89 & 91 & 4.31 & $-2.6343$ \\
			GGA+U   &       & 11.6 & 1560.89 & 89 & 4.24 & -- \\
			\hline
		\end{tabular}
	\end{table}

\subsection{Elastic properties}
\subsubsection*{Mechanical properties}
The elastic constants of a material define its mechanical behavior.
They offer crucial insights into the properties such as mechanical
stability, brittleness, ductility, rigidity, elastic moduli,
Poisson's ratio, and the elastic anisotropy of materials. The
complete elastic constant tensor was determined from calculations of
the stresses induced by slight deformations of the equilibrium
primitive cell and thus, the elastic constants $C_{ijk}$ are
determined as \cite{a28, a29}.
\begin{equation}
 C_{ijkl}=\frac{\partial\sigma_{ij}}{\partial\varepsilon_{kl}}\bigg|_{\chi}=\frac{1}{V} \frac{\partial^2 E}{\partial\varepsilon_{ij} \partial{\epsilon_{kl}}} \bigg|_\chi,
\end{equation}
where $E$  denotes the Helmholtz free energy, $\varepsilon_{ij}$ and
$\varepsilon_{kl}$ represent the imposed stress and Eulerian strain
tensors, respectively, and  $\chi$ refers to the coordinate
variables.

For the case of trigonal crystal, there are six  independent elastic
constants ($C_{11}, C_{12}, C_{13}, C_{14}, C_{33}$, and $C_{44}$)
that should satisfy the six well-known mechanical stability criteria
\cite{a30}
\begin{equation}
  C_{11}-C_{12} > 0, \quad (C_{11}+C_{12})C_{33} - 2	C_{13}^{2} > 0, \quad (C_{11}-C_{12}) C_{44}-2C_{14}^{2}  > 0.
\end{equation}
These  criteria ensure that the crystal will not undergo any
spontaneous deformation under small perturbations, 
indicating a mechanical stability. If any of these conditions are
violated, the crystal is likely to be unstable under elastic
deformations.

The elastic constants calculated for a trigonal structure R3c (161)
LaBiO$_{3}$ were  found to be  $C_{11}=216.15$~GPa, $C_{12}=127.8$~GPa,
$C_{13}=73.05$ GPa, $C_{14}=5.10$ GPa, $C_{33}=96.99$ GPa, $C_{44}=39.67$  GPa, and
$C_{66}=44.17$ GPa with GGA+U approximation that satisfy the above
mechanical stability conditions.   From the calculated elastic
constants, the mechanical parameters such as the Young modulus $(E)$, shear modulus $(G)$, Poisson's ratio $(\eta)$,  and Pugh ratio
$(r)$ are determined by using the Voigt--Reuss--Hill (VRH) average
approximation \cite{a31}.

The upper limit and lower limit of the actual effective modulus
correspond to the Voigt bound  based on uniform strain throughout a
polycrystalline assumption and the Reuss bound based on the uniform stress
throughout a polycrystalline assumption  obtained by the average
polycrystalline modulus \cite{a32}.  For trigonal lattices, Voigt
bulk modulus $(B_{V})$ and shear modulus $(G_{V})$ are
\begin{equation}
B_{V}=\frac{1}{9}[2(C_{11}+ C_{12}) + 4C_{13} + C_{33}], \quad
G_{V}=\frac{1}{15}[2C{11}-C_{12} + C_{33} - 2C_{13} + 6 C_{44} + 3(C_{11}-C_{12})/2],
\end{equation}
and the Reuss bulk modulus $(B_{R})$ and the Reuss shear modulus
$(G_{R})$ are described as:
\begin{equation}
B_{R}=\frac{(C_{11} + C_{12})C_{33}-2C_{13}^{2}}{C_{11}+ C_{12} + 2C_{33}-4C_{13}},  \quad 
G_{R}=\frac{15}{4(2S_{11} + S_{33}) -4 (S_{12}+2S_{13}) + 3(2S_{44} +S_{})},
\end{equation}
where $S_{ij}$ are the compliance constants.
The Voigt and Reuss schemes establish the theoretical maximum and
minimum limits of the real polycrystalline elastic parameters.
Therefore, the practical approximation of the bulk and shear moduli
can be considered as the arithmetic mean of the two extremes
\cite{a33}. The Hill's average for the shear modulus $(G)$ and bulk
modulus $(B)$ is described by
\begin{equation}
G=\frac{1}{2}(G_{V}+G_{R}) , \quad   B=\frac {1}{2} (B_{V}+B_{R}),
\end{equation}
where $B_{R}$  and $G_{R}$ are the Reuss bulk and shear moduli, and
$B_{V}$ and $G_{V}$ correspond to the Voigt bulk and shear moduli.
Moreover, the Young modulus $(E)$, Poisson's ratio $(\eta)$,
anisotropic index $(A^{I})$ are given by~\cite{a29, a32, a33}
\begin{equation}
E=\frac {9BG}{3B+G}, \quad \eta=\frac{3B-2G}{2(3B+G)} , \quad
A^{I}=5\frac{G_{V}}{G_{R}}+ \frac{B_{V}}{B_{R}}-6.
\end{equation}

\begin{table}[htb]
	\caption{Mechanical properties calculated from the elastic constants of
		LaBiO$_3$.}
	\label{Tab3}
%	\vspace{2ex}
	\begin{center}
		\renewcommand{\arraystretch}{1.1}
		\begin{tabular}{c c c c c c c c c c c}
			\hline
			Source & $B_V$ & $G_V$ & $B_R$ & $G_R$ & $B$ & $E$ & $G$ & $\eta$ & $r$ & $A^{I}$ \\
			\hline
			GGA+U & 119.76 & 41.81 & 92.33 & 38.12 & 106.045 & 106.52 & 39.98 & 0.33 & 0.38 & 0.78 \\
			\hline
		\end{tabular}
		\renewcommand{\arraystretch}{1}
	\end{center}
\end{table}

The calculated Bulk modulus is 109.16 GPa using the GGA+U
exchange correlation functional. The bulk modulus values derived
from elastic constants (table~\ref{Tab3}) and from the equation of state
(table~\ref{Tab2}), obtained with GGA+U (91 GPa), show a close agreement with
only minor variation.  The ductile or brittle nature of a metallic
material directly governs its mechanical behavior and likewise
dictates its mode of failure. Pugh et al. \cite{a34} introduced the
ratio of bulk modulus to shear modulus $B/G$ (2.64) as a reference
for the judgment of the ductility of a material.  For a normal
material, if its $B/G$ value exceeds 1.75, the material is ductile,
otherwise it is brittle \cite{a35}. From table~\ref{Tab3}, the calculated B/G
ratio (GGA+U) for the trigonal R3c (161) LaBiO$_{3}$ structure
indicates that the material posesses good ductile behavior.
Furthermore, Poisson's ratio $(\eta)$ serves as a mechanical
indicator that provides a valuable insight into the nature of the
bonding forces. In the evaluation of Poisson's ratio, 0.25 and 0.5
are the lower and upper limits of the central force, respectively
\cite{a35}. From table~\ref{Tab3}, the Poisson's ratio $(\eta)$ obtained
using the GGA+U approach is 0.33, which falls within the expected
bounds and signifies that the interatomic bonding forces are
predominantly central in nature. The universal anisotropic index
$(A^{I})$ is a measure to quantify the elastic anisotropic
characteristics based on the contributions of both bulk and sheared
moduli \cite{a34}. Its value equals zero for a perfectly
isotropic material. When $(A^{I}\neq 0)$, the magnitude of its
value reflects the degree of elastic anisotropy, with any positive
or negative deviation from zero indicating an increased level of
anisotropy. The computed universal anisotropic index $(A^{u})$
for trigonal structure LaBiO$_{3}$ is 0.78 with GGA+U, indicating the
anisotropic property.

\subsubsection*{Thermodynamic properties}
One of the fundamental properties of materials, the Debye temperature
$(\theta_{D})$,  correlates with many physical properties of solids
such as specific heat, elastic constant and melting temperature
\cite{a33}. One of the known approaches  to calculate the Debye
temperature ($\theta_D$) can be estimated from the averaged sound velocity
$(C_{m})$ and is given by \cite{a35} 

\begin{equation}
\theta_D=\frac{h}{k_{\rm B}}
\bigg[ \frac{3n}{4\piup} \bigg(\frac{N_{\rm A}\rho}{M}\bigg)
\bigg]^{\frac{1}{3}} C_m ,
\end{equation}
where $h$ is Plank's constant, $k_{\rm B}$ is Boltzmann's constant, $N_{\rm A}$
is Avogadro's number, $\rho$  is density, $M$ is molecular weight
and $n$ is the number of atoms in a formula unit. Then, $C_m$  is
\begin{equation}
C_m= \bigg[ \frac{1}{3} \bigg( \frac{2}{c_t^3} + \frac{1}{c_l^3}
\bigg) \bigg]^{-\frac{1}{3}},
\end{equation}
where $c_l$  and $c_t$ are the longitudinal and transverse sound velocity. These velocities can be obtained from density, shear and bulk modulus of the material as:
\begin{align}
 &c_l= \bigg(\frac{B+\frac{4}{3}G}{\rho}\bigg)^\frac{1}{2},\\
 &c_t= \bigg( \frac{G}{\rho}\bigg)^\frac{1}{2}.
\end{align}
Using a relaxed unit cell volume of 1042.064~a.u.$^3$ ($1.544\times 10^{-22}$~cm$^3$) from the Quantum ESPRESSO output, the calculated mass density of BaTiO$_3$ is 8.54  g$\cdot$cm$^{-3}$  ($8.54\times 10^{3}$~kg$\cdot$m$^{-3}$). This value was derived using the standard relation between unit-cell mass and volume, which incorporates the atomic masses of Ba, Ti, and O.

Based on the determined elastic constants and density, the derived longitudinal
$c_l$  and transverse $c_t$  sound velocities, the average sound
velocity $C_m$, and the Debye temperature $\theta_D$ for  LaBiO$_3$
are presented in table~\ref{Tab4}.

\begin{table}[htb]
\caption{Computed parameters of sound velocities, and Debye
temperature of  LaBiO$_3$.} \label{Tab4} %\vspace{1ex}
\begin{center}
\renewcommand{\arraystretch}{1.1}
\begin{tabular}{ c  c c c c }
\hline
      Sourse &  $c_l$ (m/s) &  $ c_t$ (m/s)  &  $C_m $ (m/s) &  $\theta_D$ (K)   \\
\hline
Calculated value (GGA+U)  & $4322.63$ & $2171.10$ &  $2405.32$ & $288.82$      \\
\hline

\end{tabular}
\renewcommand{\arraystretch}{1}
\end{center}
\end{table}

\begin{figure}[h]
	\centerline{\includegraphics[scale=0.6]{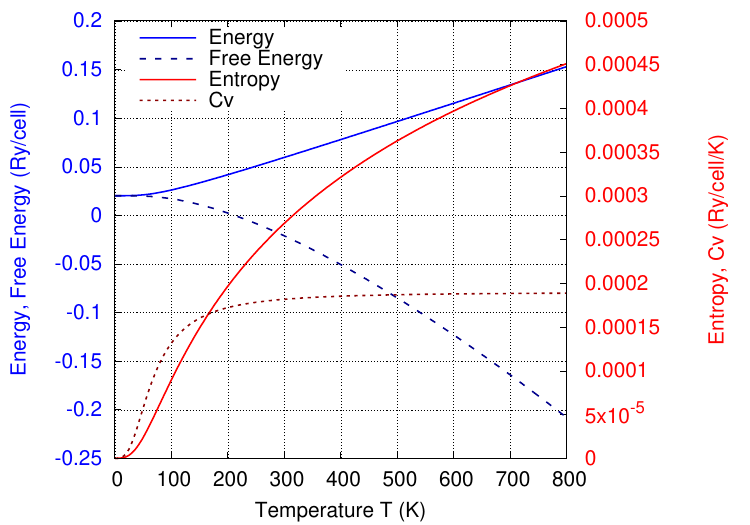}}
	\caption{(Colour online) Debye vibrational energy, free energy,  heat capacity, and
		entropy.} \label{fig-3}
\end{figure}

The thermodynamic properties of the perovskite structure were
further examined using the quasi-harmonic Debye model \cite{a38}. In
this framework, the nonequilibrium Gibbs free energy $G(V, P, T)$ is
described by:
\begin{equation}
  G(V;P,T)=E(V)+PV+F_{vib} [\theta_{D}(V);T].
\end{equation}
Here, $E(V)$ represents the total energy per unit cell, $PV$ stands
for the applied hydrostatic pressure, $\theta(V)$ denotes the Debye
temperature as a function of volume, and $F_{vib}$ corresponds to
the vibrational Helmholtz free energy. Within the quasi-harmonic
Debye model, which incorporates the phonon density of states, the
vibrational Helmholtz free energy $F_{vib}$ is expressed as
\cite{a38, a39}:
\begin{equation}
	F_{vib}(\theta_{D};T)=nk_{\rm B} T\left[ \frac{9\theta}{8T} + 3\ln\left(1-\re^{-{\theta}/{T}}\right) -D\bigg(\frac{\theta_{D}}{T}\bigg)\right],
\end{equation}
where $n$ is the number of atoms per formula unit, %$k_{\rm B}$ is Boltzmann's constant, 
 $D(\theta_D/T)$  denotes the Debye integral.
For an isotropic solid, $\theta_{D}$ can be expressed as \cite{a30,
a32}
\begin{equation}
\theta_D=\frac{\hbar}{k_{\rm B}}  (6\piup^{2} V^{1/2} n)^{1/3} f(\sigma)
\sqrt{B_s/M}.
\end{equation}
Here, $M$ is the molecular mass per unit cell and $B_s$ is the
adiabatic bulk modulus, which characterizes the compressibility of
the crystal and is approximated by the static compressibility as
\cite{a30}:
\begin{equation}
B_s\cong B(V)=V \frac{\rd^{2} E(V)}{\rd V^{2}} .
\end{equation}
Heat capacity at constant volume $C_v$, and the entropy are
expressed by
\begin{align}
 &C_{v}=3nk_{\rm B}\left[4D\left(\frac{\theta_D}{T}\right)-\frac {3 \theta_D/T}{\re^{\theta_D }-1}\right],\\
&S=nk_{\rm B}\left[4D\left(\frac{\theta}{T}\right)-3\ln\left(1-\re^{-\frac{\theta}{T}}\right)\right].
\end{align}
We evaluated the thermodynamic properties of LaBiO$_3$ including
enthalpy, Gibbs free energy, entropy, and heat capacity over the
temperature range of 0--800 K, as presented in figure~\ref{fig-3}. For the
analysis, both volume and temperature were considered as independent
variables. As demonstrated in figure~\ref{fig-3}, the entropy and
heat capacity approach zero at 0 K, which is consistent with the third law of
thermodynamics. As temperature rises, the Gibbs free energy
decreases gradually, whereas the entropy increases sharply. These
trends are consistent with previously reported results in 
\cite{a32, a33}. Consequently, the enthalpy rises nearly linearly
with temperature. At higher temperatures, this increment in enthalpy
contributes to a reduction in the defect-related free energy. The
heat capacity increases sharply below 200 K and then approaches
the classical Dulong--Petit limit 500~K$^{-1}$mol$^{-1}$ above 200~K.
This trend is consistent with anharmonic corrections within the
Debye model, as reported in \cite{a35, a36}. Moreover, the
smooth and continuous variation in heat capacity indicates the
absence of any phase transitions up to 800 K, in agreement with
reference~\cite{a30}. Overall, these results provide useful insights for
future studies.

\subsection{Electronic parameters: band structure and density of states}
\begin{figure} [h]
	\centering
	\includegraphics[width=0.45\textwidth]{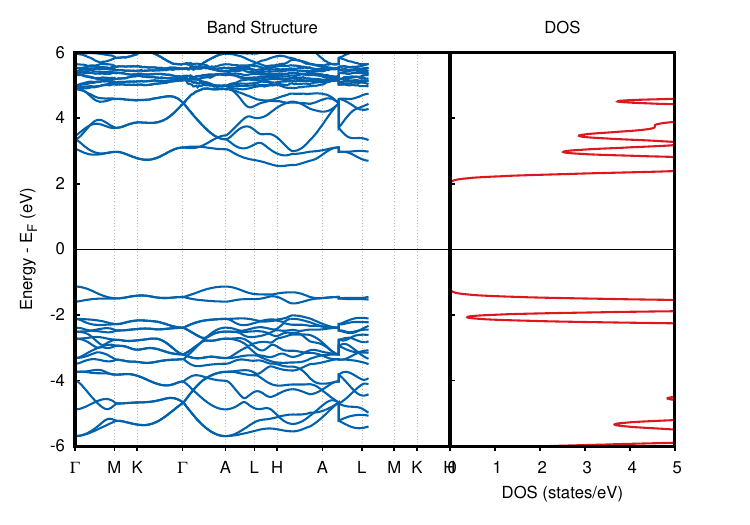}
	%	\hfill
	\includegraphics[width=0.45\textwidth]{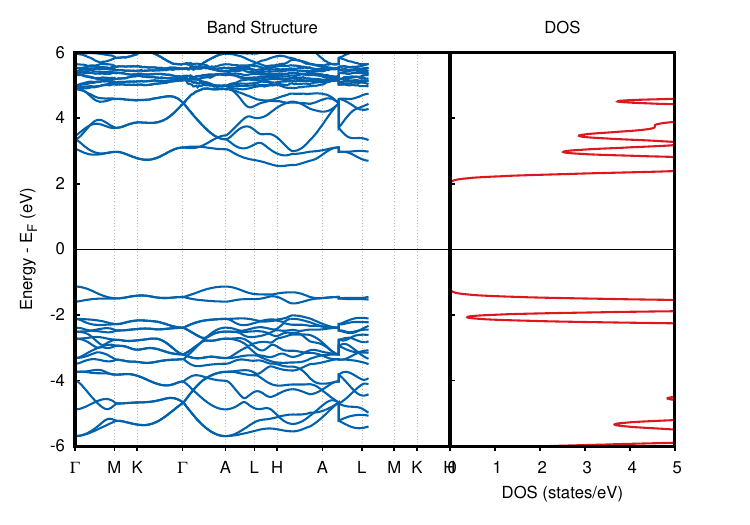}
	\caption{(Colour online) Band structure and density state of LaBiO$_3$ with respect to GGA-PBE (left-hand) and GGA+U (right-hand).}%
	\label{Fig-4}
\end{figure}

The density of states and electronic band structure typically offer
an appropriate information for a comprehensive characterization of electronic properties of a material. In this work, the electronic band
structure and density states of LaBiO$_3$ along the high-symmetry
directions of the Brillouin zone are calculated using the GGA and
GGA+U approach for the exchange-correlation potential, as
illustrated in figure~\ref{Fig-4}. The calculated band structure
shows that the trigonal phase of LaBiO$_3$ is an indirect large
band gap material. Using the GGA and GGA+U exchange-correlation
potential, the computed energy gap values for trigonal LaBiO$_3$ are 3.48 eV
and 3.54 eV, respectively. The band gap value obtained by GGA+U for
exchange correlation potential is somewhat larger than the value
calculated using GGA approximation.

\subsection{Optical properties}
\begin{figure}[h!]
	\centerline{\includegraphics[scale=0.75]{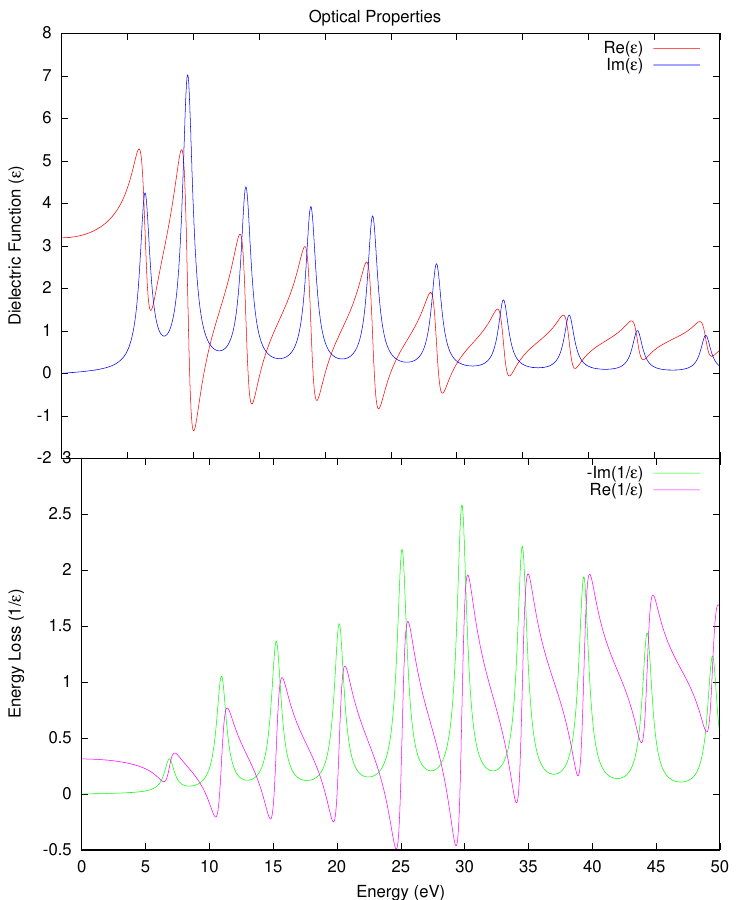}}
	\caption{(Colour online) The real and imaginary part of macroscopic dielectric
		function, and the electron energy loss function as function of
		energy.} \label{fig-5}
\end{figure}

To compute the optical properties, the complex dielectric function
$\varepsilon (\omega)$ is employed within an adiabatic exchange
correlation functionals (AXCA) \cite{a40}. The imaginary component,
$\varepsilon_{2} (\omega)$, is determined from the  dipole
transition matrix elements between valence and conduction band
states, employing the long-wavelength approximation. The dispersion
of the real part of the dielectric function $\varepsilon_{1}
(\omega)$ was determined from the imaginary part using the Kramers-Kronig
relations. Furthermore, the experimentally accessible  optical
parameters such as the refractive index  $\Re[\varepsilon_M
(\omega)]_{q\rightarrow 0}$, the absorption coefficient
$\Im[\varepsilon_M (\omega)]_{q\rightarrow 0}$, and the energy loss
function $-\Im \left[ {1}/{\varepsilon_M
(\omega)}\right]_{q\rightarrow 0}$, are explored in trigonal
structure of LaBiO$_3$. The calculated optical properties are
demonstrated in figure~\ref{fig-5}.

In figure~\ref{fig-5}, the absorption coefficient for LaBiO$_3$ shows a
maximum peak at $10$ eV, attributed to single excitation. By
contrast, the peak in the electron energy loss function (EELF),
associated with collective plasmon excitation, occurs at a higher
energy of 30 eV. The maximum of the electron energy loss function
is observed when $\Re\varepsilon_M (\omega)=0$  and
$\Im\varepsilon_M (\omega)$ is small. The imaginary component
$\varepsilon_{2} (\omega)$ of the dielectric function governs the
optical transition processes, with each peak representing a specific
electronic transition.
%\begin{figure}[htb]
%\centerline{\includegraphics[width=0.65\textwidth]{bands_dos_hub}}
%\caption{Absorption coefficient of LaBiO$_3$ with no interaction and
%including electron interactions.} \label{fig-6}
%\end{figure}

\begin{figure}[h!]
	\centerline{\includegraphics[scale=0.75]{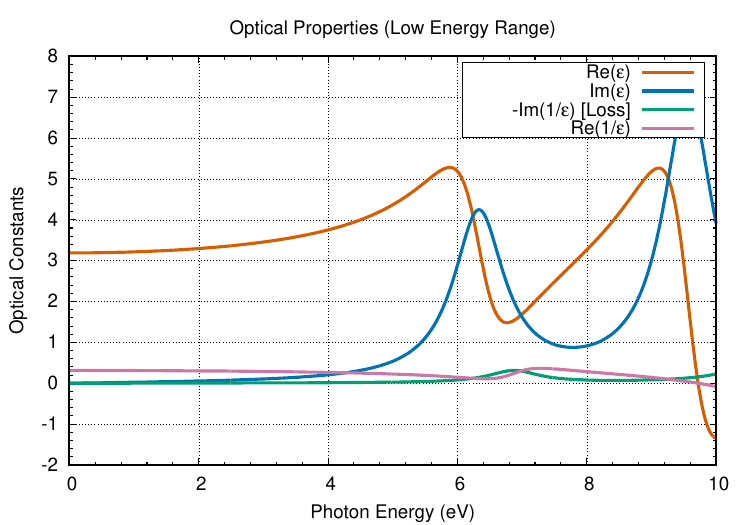}}
	\caption{(Colour online) Absorption coefficient of LaBiO$_3$ at low energy regime}
	\label{fig-6}
\end{figure}

At low energy range (0--10 eV), the optical properties  clearly
describe the system as shown in figure~\ref{fig-6}. The real part 
$\Re\varepsilon_M (\omega)$ describes the  refractive index of the material 
and how much it slows down the light. The anomalous dispersion is
observed around 6 eV and 9 eV.  At the far right (10 eV), the value
drops below zero which usually indicates the onset of plasma-like
behavior. The imaginary part $\Im\varepsilon_M (\omega)$ represents
the energy absorption and LaBiO$_3$ is transparent until about 3.6
eV due to its wide band gap. The peaks at 6.2 eV and 9.5 eV
correspond to specific electronic transitions from valence band to
conduction band.

The electron energy loss function $-\Im \left[ {1}/{\varepsilon_M(\omega)}\right]$ describes how much energy an
electron loses as it passes through the bulk of the material. Peaks
in this line (10 eV) typically identify plasmone. How the absorption
constant and a refractive index changes with respect to photon enegy
is described in table~\ref{Tab 5}.


\begin{table}[h]
\caption{Calculated values for the real and imaginary parts of the dielectric function and the refractive index at selected photon energies
	LaBiO$_3$.} \label{Tab 5} \vspace{1ex}
\begin{center}
	\renewcommand{\arraystretch}{1.1}
	\begin{tabular}{c c c c}
		\hline
		Photon energy (E) &  $\Re\varepsilon_M (\omega)$ & $\Im\varepsilon_M (\omega)$  & Calculated refractive index $n(\omega)$  \\
		\hline
		2.0 eV (visible range)&   3.3  & 0.0 & 1.82    \\
		\hline
		4.0 eV (near UV)  &   3.8 & 0.3  & 1.95\\
		\hline
		6.0 eV (resonance peak)& 5.3   & 3.5 & 2.8   \\
		\hline
		10.0 eV (high energy) & $-1.4$   & 3.8  & 1.15\\
		\hline
	\end{tabular}
	\renewcommand{\arraystretch}{1}
\end{center}
\end{table}

\begin{figure}[h]
	\centerline{\includegraphics[scale=0.7]{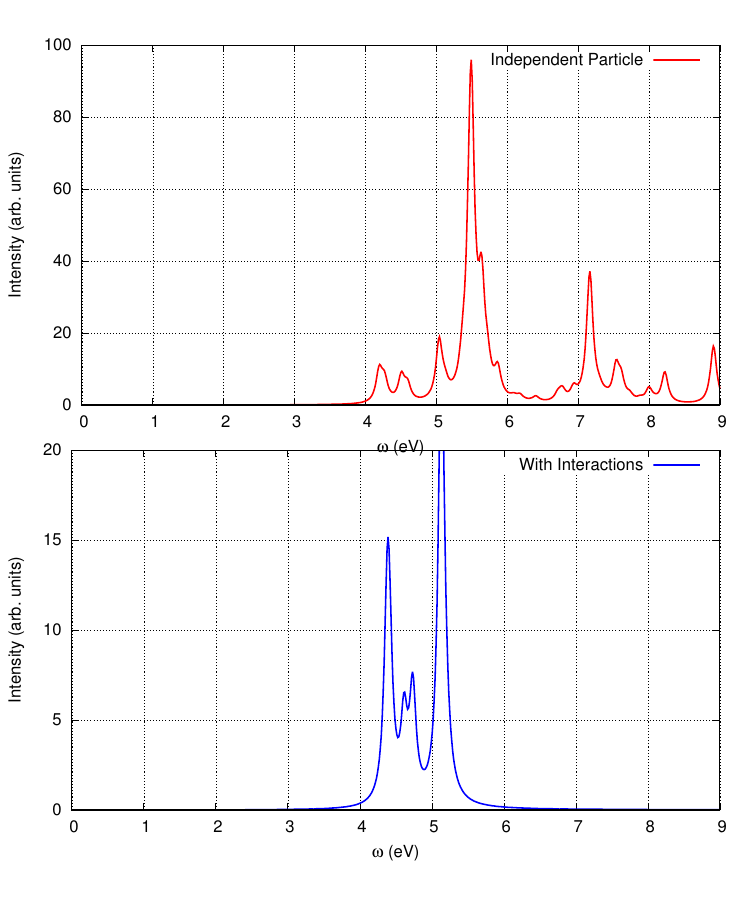}}
	\caption{(Colour online) Absorption coefficient of LaBiO$_3$ with no interaction and including electron interactions.}
	\label{fig-7}
\end{figure}

Moreover, the absorption coefficient of trigonal LaBiO$_3$ is
determined using independent particle approximation (without
considering electron interaction) and including electron
interactions (Hartree and exchange correlation effects) as shown in
figure~\ref{fig-7}. The red shift of the peak is observed in the
absorption coefficient when electron-electron interaction is
considered. Moreover, the absorption spectra are quenched due to the
local field effects.

\subsection{Transport properties}

\begin{figure}[htb]
\centerline{\includegraphics[scale=0.65]{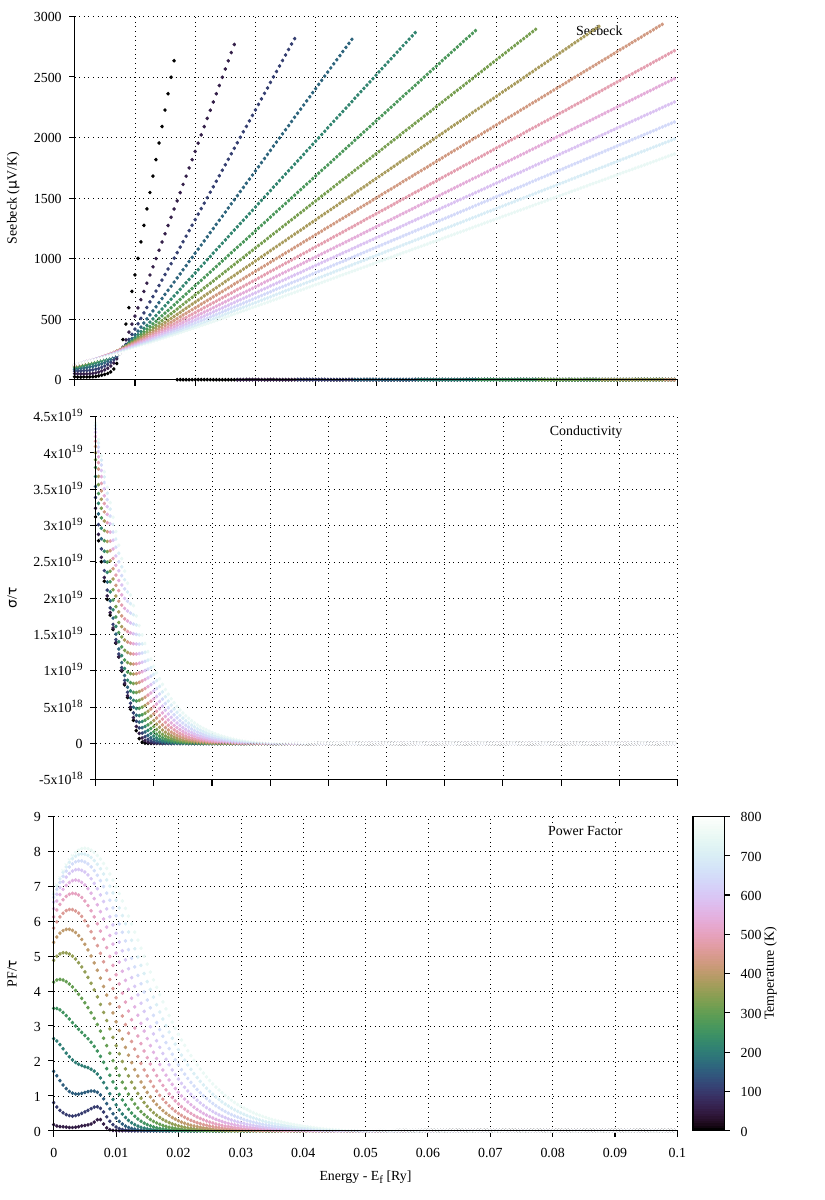}}
\caption{(Colour online) [Top] Power factor per relaxation time ($10^{11}$~W/m$\cdot$K$^{2}$$\cdot$s), [middle] conductivity per relaxation time ($10^{19}/\Omega$$\cdot$cm$\cdot$s), and [bottom] Seebeck (\textmu V/K)  as a function of energy $[E-E_F]$ (Ry) for different temperatures 50--800 K with step of 50 K.} \label{fig-8}
\end{figure}

Boltzmann transport theory based on the first-principles electronic structure
calculations was employed to determine the semi-classical transport coefficients. In this transport property investigation, only 
the electronic contribution is considered. This approach helps to evaluate the
temperature dependence of the power factor (PF), electrical
conductivity per relaxation time $(\sigma/\tau)$, and Seebeck
coefficient $(S)$. The resulting transport coefficients for
LaBiO$_3$ are presented in figure~\ref{fig-8} .

In figure~\ref{fig-8} (top) the Seebeck coefficient increases
considerably as Fermi energy is shifted towards the $p$-type region.
Moreover, its magnitude increases with an increment of temperature.
Particularly, strong thermopower  especially in the $p$-type side
indicates that LaBiO$_3$ has favorable thermoelectric behavior.

\begin{figure}[h]
\centerline{\includegraphics[scale=0.83]{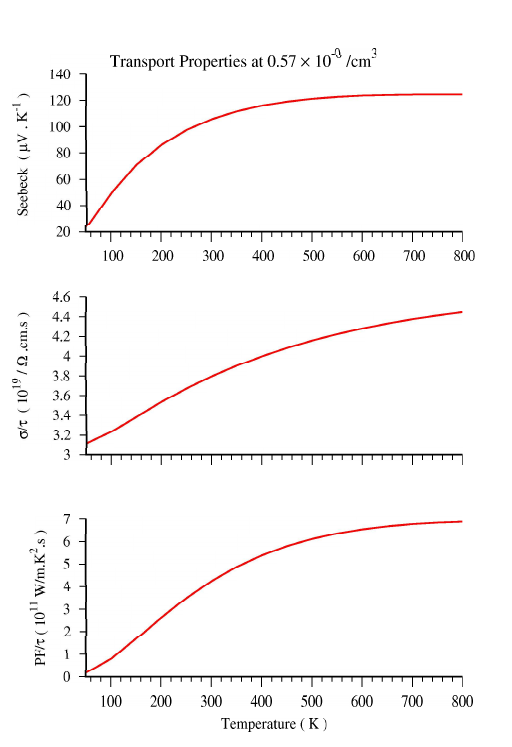}}
\caption{(Colour online) [Top] Seebeck (\textmu V/K), [middle] conductivity per
relaxation time $(10^{19}/\Omega\cdot$cm$\cdot$s), and [bottom] power factor
per relaxation time $(10^{11}$~W/m$\cdot$K$^{2}$$\cdot$s) as a function of
temperature.} \label{fig-9}
\end{figure}

Figure~\ref{fig-8}  (middle) describes a strong peak of electrical
conductivity per relaxation time near the Fermi level $(E-E_{F} =
0)$, and then rapidly drops. This is attributed to the high density of 
states near the Fermi level, which enhances the carrier concentration and, 
consequently, the conductivity. As the carrier density decreases further 
from the Fermi energy, the conductivity per relaxation time $({\sigma}/{\tau})$ 
drops accordingly. The result indicates that LaBiO$_3$ becomes
more conducting near the intrinsic Fermi level. Figure~\ref{fig-8}
(bottom) reveals that the PF initially rises as the Fermi level moves
away from the reference point (towards $p$-type region). It reaches a
maximum for each temperature and then sharply decreases. Higher
temperatures ($\geqslant 600$ K) show a larger peak of PF, which indicates a better
thermoelectric performance at higher temperature. Moreover, at a specific temperature 
of $600$~K, the estimated relaxation time $\tau$ is $10^{-14}$~s, while the 
computed maximum conductivity per relaxation time ${\sigma}/{\tau}$ is
 $4.3 \times 10^{19}$~($\Omega \cdot$cm$\cdot$s)$^{-1}$. By combining 
 these parameters, the resulting electrical conductivity $\sigma$ is determined to 
 be $4.3 \times 10^{5}$~($\Omega\cdot$cm)$^{-1}$, a value that indicates 
 high conductivity for the material under these specific thermal conditions.  Thus, LaBiO$_3$ shows
promise for thermoelectric applications especially in $p$-type regime
at elevated temperatures. The transport coefficients as a function of temperature are
described in figure~\ref{fig-9}.

The transport properties indicate that the material is a $p$-type
semiconductor, as evidenced by the positive Seebeck coefficient which
rises from $20$~\textmu{V/K} and plateaus near $125$~\textmu{V/K} at higher temperatures. Both the scaled electrical
conductivity ($\sigma/\tau$) and the scaled power factor ($PF/\tau$)
exhibit a steady monotonous increase across the temperature range
from 50 K to 800 K, suggesting that the electronic
performance of the material is optimized for high-temperature applications.
Specifically, the sharp upward trajectory of the power factor, driven by
the squared relationship with the Seebeck coefficient ($S^2$),
indicates that the material reaches its maximum thermoelectric
efficiency at approximately 800 K, where it reaches its peak value
of $7 \times 10^{11}$~W/(m$\cdot$K$^{2}$$\cdot$s).

In conclusion, LaBiO$_3$ demonstrates a promising thermoelectric
performance, particularly under $p$-type doping and high temperatures
($\geqslant 500$ K). Enhanced power factor and Seebeck coefficient at
elevated temperatures indicate suitability for waste heat energy
conversion and thermoelectric devices.

\section{Conclusion}
The structural, electronic, elastic, thermodynamic, optical and
thermoelectric parameters of trigonal LaBiO$_3$ perovskite were
investigated via first-principles calculations employing the Quantum
Espresso package. Using the GGA-PBE functional, the lattice
parameter was computed as 11.3 Bohr, while the GGA+U method yielded
11.6 Bohr. The formation energy is determined and its value is
$-2.6343$~eV/atom (negative values) indicating that  the energy gained
during the formation confirms that the compounds are stable. The elastic
constants of LaBiO$_3$ were determined and employed to evaluate
the mechanical properties, including the bulk modulus, shear modulus,
Young's modulus, and elastic anisotropy. Using GGA+U, the Poisson's
ratio is found to be 0.33, suggesting that the interatomic forces
are predominantly central. The computed $B/ G$ ratio of 2.7 using
GGA+U confirms that LaBiO$_3$ exhibits a ductile behavior. The band
gap value was calculated  using GGA+U approximation for the exchange
correlation potential. The band gap obtained using the
Hubbard-corrected GGA+U method for on-site interactions is 3.51~eV.
Optical properties, including the absorption coefficient, refractive
index, and electron energy loss function, were calculated and
analyzed in detail. The Hartree and exchange correlation effect in
absorption spectra results in blue shift of the peak. Moreover, the
absorption spectra are quenched  due to the local field effect when
the electron-electron interaction is considered.  Finally,  the
semi-classical transport coefficients were calculated, and they
demonstrate that LaBiO$_3$  may be suitable for waste heat energy
conversion and thermoelectric devices. It can also be used for
photocatalysis, for energy conversion and storage.
\\

\textbf{Funding:} This study was supported by a staff research grant from the College of Natural Sciences,
Jimma University. 

\section*{Acknowledgements}
The authors would like to acknowledge CHPC LENGAU South Africa for
High Performance Cluster Machine (HPC) support during the
computation.

%\bibliographystyle{cmpj}
%\bibliography{cmpjxampl}

\ukrainianpart

\title{Дослідження перовскіту LaBiO$_{3}$ для фотоелектричних, термоелектричних та оптоелектронних застосувань}

\author{M. М. Волдемаріам, 
	Н. Г. Дебело, 
	Т.~К.~Туфа, 
	E.~M.~Гурмеса, 
	С.~Х.~Діду, 
	С.~Н.~Асфав, 
	Д.~Т.~Кено}

\address{
	Фізичний факультет університету Джимми, п/с 378, Джимма, Ефіопія
}

\makeukrtitle

\begin{abstract}
	\tolerance=3000%
	Використовуючи теорію функціоналу густини (DFT) з псевдопотенціалом ONCVVPSP та функціоналом PBE, досліджуються структурні, електронні, пружні, оптичні та термоелектричні властивості тригонального перовскітного оксиду LaBiO$_{3}$ (просторова група R3c).
	За допомогою рівняння стану були визначені параметри основного стану: стала ґратки, об'єм, модуль об'ємної пружності та його похідна за тиском. Застосування поправки Хаббарда (GGA+U) виявило непряму широку заборонену зону 3.51 еВ. Механічні властивості, включаючи коефіцієнт анізотропії, модуль пружності та коефіцієнт Пуассона, були розраховані за схемою усереднення Фойгта-Ройса-Хілла. 
	Відношення модуля об'ємної пружності до модуля зсуву ідентифікує тригональну фазу як пластичну. Крім того, були обчислені температури Дебая та швидкості звуку. Оптичні характеристики (коефіцієнт поглинання, показник заломлення та функція втрат енергії електронів) були оцінені в спектральному діапазоні 0--35 еВ. Для оцінки термоелектричного потенціалу матеріалу були розраховані напівкласичні коефіцієнти переносу, включаючи електропровідність, коефіцієнт Зеєбека та коефіцієнт потужності.
	\keywords теорія функціоналу густини, LaBiO$_{3}$, електронні властивості, пружні властивості, оптичні властивості, транспортні властивості
\end{abstract}

\lastpage
\end{document}